\documentclass[preprint,nofootinbib,showpacs,superscriptaddress,aps,prc]{revtex4-2}
\usepackage{graphicx}
\usepackage{amstext}
\usepackage{amssymb}
\usepackage[usenames]{color}

\begin{document}

\title{Space-time structure of particle emission and femtoscopy scales in ultrarelativistic heavy-ion collisions}

\author{Yu.~M.~Sinyukov}
\affiliation{Department of High-Density Energy Physics, Bogolyubov Institute for Theoretical Physics, 14b Metrolohichna street, 03143 Kyiv, Ukraine}
\affiliation{Faculty of Physics, Warsaw University of Technology, 75 Koszykowa street, 00-662, Warsaw, Poland}
\author{V.~M.~Shapoval} 
\affiliation{Department of High-Density Energy Physics, Bogolyubov Institute for Theoretical Physics, 14b Metrolohichna street, 03143 Kyiv, Ukraine}
\author{M.~D.~Adzhymambetov}
\affiliation{Department of High-Density Energy Physics, Bogolyubov Institute for Theoretical Physics, 14b Metrolohichna street, 03143 Kyiv, Ukraine}

\begin{abstract}
The analysis of the spatiotemporal picture of particle radiation in relativistic heavy-ion
collisions in terms of correlation femtoscopy scales, emission and source functions 
allows one to probe the character of evolution of the system created in the collision. 
Realistic models, like the integrated hydrokinetic model (iHKM), used in
the present work, are able to simulate the entire evolution process of strongly interacting matter produced
in high-energy nuclear collision. The mentioned model describes all the stages of the system's evolution,
including formation of the very initial state and its consequent gradual thermalization,
hydrodynamic expansion and afterburner hadronic cascade, that can help researchers to figure out 
the specific details of the process and better understand the formation mechanisms of certain observables. 
In the current paper we investigate the behavior of the pion and kaon interferometry radii and their 
connection with emission functions
in ultrarelativistic heavy-ion collisions at the Large Hadron Collider within iHKM. We are focusing
on the study of the emission time scales at different energies for both particle species (pions and kaons) aiming
to get deeper insight into relation of these scales and the peculiarities of the mentioned system's collective
expansion and decay with the experimentally observed femtoscopy radii. 
One of our main interests is the problem of the total system's lifetime estimation based on the femtoscopy analysis.
\end{abstract}

\maketitle

Keywords: kaon; pion; femtoscopy radius; emission function; emission time; particlization.

\section{Introduction}

The investigation of the evolution process character for the extremely small, hot, dense and 
rapidly expanding systems created in relativistic heavy-ion collisions was always of great interest 
to the researchers, since the first studies of these phenomena started. The comprehensive analysis
demonstrates that soon after the smashing of the two colliding nuclei, the strongly coupled quark-gluon system is formed. Being at first non-equlibrium, this system
some time later gets thermalized and continues expanding as a whole liquid-like piece of matter~\cite{qgp1,qgp2,qgp3,qgp4}.
Its expansion at this stage can be described within a relativistic hydrodynamics approach.
As the system expands, it gradually cools down, gets dedensified and begins losing equilibrium.
Below some temperature or energy density (usually the temperatures close to 150-160~MeV are spoken of)
the system disintegrates, and its parts one by one are transformed into sets of hadrons.
However, the produced hadrons do not likely become free immediately after the fireball decay. 
The latter would imply a sudden switching from a near-zero mean free path length (in hydrodynamic medium) to a near-infinity one (in free streaming regime), which seems hardly feasible. 
The particles would rather continue interacting intensively and evolving collectively, 
as strongly interacting gas for some time.
So, to estimate the overall lifetime of the system created after the collision, one should in theory
account for the hadronic phase of the matter evolution and try to find out the time scale (at least approximate),
during which the hadronic medium stays connected enough to be considered a unified system.

In the literature, however, there are two different approaches to the stated problem. 
The first one supposes that right after the hadronization (and maybe a very short hadronic stage)
the so-called ``freeze-out'' takes place, so that the chemical composition of the system and,
almost simultaneously, the particle momentum spectra are frozen and do not change anymore,
except for decays of resonances.
Such an assumption goes back to the pioneer works by Fermi, Pomeranchuk, and Landau~\cite{fermi,pomeranchuk,landau}
and is typical for thermal models with statistical
hadronization~\cite{therm1,therm2,therm3,therm4,therm5}, hydro-inspired freeze-out parametrizations, e.g. 
Cracow~\cite{cracow1,cracow2}, Buda-Lund~\cite{buda-lund} and Blast Wave~\cite{bw1,bw2,bw3}, 
and hydrodynamics models with statistical
hadronization~\cite{hydrostat1,hydrostat2,ollitrault,shuryak1,heinz,schlei,sollfrank,shuryak2,
nonaka,hirano1,hirano2,heinz2010,bozek,kisiel}.
The latter usually employ the Cooper-Frye formalism~\cite{cooper-frye} or its 
generalizations~\cite{molnar} to hadronize the system at the freeze-out hypersurface. 
There are various computer codes, developed for calculations of
statistical hadronization in different models, e.g. SHARE~\cite{share1,share2}, 
THERMUS~\cite{thermus}, used for the extraction of thermodynamic parameters 
from the fits to ratios of particle yields, or THERMINATOR~\cite{therminator1,therminator2}, used 
as event generator, producing particles based on the thermodynamic parameter values 
specified at certain freeze-out hypersurface. 

The second approach admits the existence of post-hydro ``afterburner'' stage of the collision
lasting about 5-10~fm/$c$ (depending on the type of the utilized cascade model), or until the 
temperature drops to 80-120~MeV. This stage may play an important
role in the measured observables formation. 
The corresponding theoretical description of heavy-ion collisions
in this case is carried out within various hybrid (hydro+cascade) and full evolutionary models, aiming to simulate 
the entire process of the created system's evolution, from the initial state formation to the final hadron cascade 
stage~\cite{hybrid1,hybrid2,hybrid3,hybrid35,hybrid4,hybrid5,hybrid6,hybrid7,hybrid8,hybrid9,hybrid10,hybrid11}.
The study presented in this paper is carried out in the most developed complete model --- integrated 
HydroKinetic Model (iHKM), which includes as one of the most important stages of the system's 
evolution, also a description of the process of thermalization/hydrodynamization of initially non-
thermal matter~\cite{ihkm2016}.

The theoretical advantage of the second class of models in view of this article's topic is obvious: 
if the post-hydrodynamics stage 
gives nearly zero contribution to observables, this can be proved just and only within the 
``afterburner'' class of approaches, accounting for the real cross-sections of particle interactions 
at the final stage of the system's evolution. 

The existing experimental data on $p_T$ spectra
do not allow to give clear preference to one of these approaches and yield the estimates 
for kinetic freeze-out
temperature in a very wide range (like $100-140$~MeV~\cite{hirano2}), 
depending on the particle species selected for the analysis,
momentum cuts, inclusion/exclusion of resonance decays and so on. 
The results on $K^*$ resonance yields
and $K^*/K$ yields ratios~\cite{kstexp1,kstexp2,alicekst} 
speak in favor of the afterburner scenario --- the experimental 
$K^*/K$ ratio values decrease when one goes from peripheral to central collisions,
in contrast with almost flat dependency (keeping close to the measured values for peripheral 
events) obtained in the models which do not account for the hadron re-scattering stage of 
collision (e.g., thermal model, blast-wave model). Such models also overestimate the measured 
$K^*$ yield values. 
The observed suppression of the short-lived $K^*$ yield ratio to kaon yield is likely
due to re-scattering of $K^*$ decay products during the afterburner collision stage~\footnote{Such a 
hypothesis is additionally supported by a nearly flat $\phi(1020)/K$ ratio dependency on centrality, 
observed in the experiment. This data feature agrees with the assumption that due to a large lifetime of the $\phi$ resonance, 
its decay products should not noticeably interact with the dense hadronic medium at the post-hydro stage.}, 
and this effect is stronger in central events because of higher number of produced particles
and longer hadronic stage duration in this case. 
Note, however, that the system's lifetime determined
based on these results depends on the cross-sections and reaction channels 
used in the model for the hadronic stage.

The correlation femtoscopy data can also be considered a potential source of information 
on the duration of the matter collective motion in high-energy A+A collisions.
For instance, in the ALICE Collaboration's experimental analysis~\cite{alice-femto} the ``freeze-out time''
value for the $2.76$~TeV Pb+Pb collisions at the LHC is defined from the fit to the measured 
longitudinal femtoscopy radii dependency on $m_T$ (the result is 10-11~fm/$c$), but the fitting 
formula does not account for the strong transverse collective flow that should develop in such 
collisions.
In our previous papers \cite{mtscale276, rhic-ihkm, mtscale502} a simple method for the estimation of the times
of pion and kaon maximal emission, based on the particle $p_T$ spectra and the dependencies 
of longitudinal femtoscopy radii $R_{long}$ on the pair transverse mass $m_T$,  
was proposed and tested in simulations within the different versions of the (integrated) hydrokinetic 
model~\cite{hkm,ihkm2016}. The method accounts even for intensive transverse flow. 
Soon after the first of the mentioned publications,
the proposed prescription was used by the ALICE Collaboration in their experimental analysis
of kaon femtoscopy for the $2.76$~TeV Pb+Pb collisions at the LHC \cite{alice-mt}. It provided 
estimates for the kinetic freeze-out temperature, $T_{kin}=144$~MeV, and the maximal
emission times, $\tau_{\pi}=9.5\pm0.2$~fm/$c$ and $\tau_{K}=11.6\pm0.1$~fm/$c$, 
close to those obtained in our paper~\cite{mtscale276} (see the details below).
 
A clear influence of the post-hydrodynamic phase on the soft physics results is also noted in our analysis of
different particle number ratios~\cite{ratios,lhc502-ihkm} at the LHC energies. 
These studies include comparison 
of the values calculated in full iHKM simulation regime with those obtained in a reduced regime 
without a hadronic cascade stage and also with the thermal model results~\cite{therm1,therm2}. 

According to simulation analysis presented in~\cite{kstar1,kstar2} 
a large fraction (about 60\% or even more) of
identified $K^*(892)$ resonances at the LHC are produced at the hadronic phase of collision,
while primary particles coming from the QGP hadronization cannot be observed due to rescattering of their
decay daughters. By contrast, primary $\phi(1020)$ mesons with lifetime $\approx 50$~fm/$c$ can be identified
without problems, apparently because they decay after the finishing of the intensive hadron rescattering phase
(however, iHKM simulation results in~\cite{kstar1} do not exclude that the hadronic stage may lead to
an additional increase in identified $\phi(1020)$ number due to $KK$ correlation effect).
The recent ALICE Collaboration data on $K^*/K$ and $\phi/K$ yield ratios in $5.02$~TeV
Pb+Pb collisions at the LHC confirm ``the dominance of the rescattering effect in the hadronic phase''~\cite{alicekst} for $K^*$ results and its small significance in the $\phi$ case.  

All these results seem to support
the conception of continuous (rather than sudden) freeze-out and prolonged afterburner stage.
In accordance with this evolution picture, particles possesing different momenta and originating from
different sources (primary particles and those leaving the system after elastic/inelastic reactions
with the hadronic medium at the afterburner stage of collision) can be expected to
radiate from the system at different times.

In the current paper we try to investigate the process of pion and kaon emission considered 
in References~\cite{mtscale276, mtscale502} in more detail, clarify the relationship between the extracted
emission times and the femtoscopy radii, and, if possible, propose a way for estimation of the full system's 
lifetime, including the afterburner stage, based on our analysis of ultrarelativistic heavy-ion collisions 
within the integrated hydrokinetic model.

\section{Research motivation}

In our studies \cite{mtscale276, rhic-ihkm, mtscale502} the transverse momentum spectra 
and femtoscopy radii for the A+A collisions at the top RHIC and the two LHC energies
were fitted with simple analytical formulas, containing parameters describing the system's effective temperature, 
the strength of transverse collective flow and the time of maximal emission for particles of each sort. The fitting formulas 
were obtained in approximation, suggesting that all the particles of a given species with 
comparably low momenta ($p_T \lesssim 0.5$~GeV/$c$) are emitted from the system at 
the corresponding hypersurface of constant proper time $\tau_\mathrm{m.e.}$, limited in transverse direction. Of course, the real emission picture for both
pion and kaon mesons is much more complicated, so that the extracted maximal emission times correspond only
to some effective values, approximately reflecting the time scales when the emission process 
for a certain particle sort is the most intensive.

To cross-check the fitting results and to obtain a more detailed picture of the particle emission process
in \cite{mtscale276, rhic-ihkm, mtscale502} we additionally built plots of averaged emission functions
in coordinates $(r_T,\tau)$ for particles with small $p_T$ (in accordance with the earlier applied approximation).
The figures showed that the emission function maximum for pions should be close to the particlization time
for the center of the expanding system, that was in agreement with the previously obtained fitting result.
As for the kaon emission function, it had two apparent maxima (one close to the pionic one and the other
lying about 4~fm/$c$ higher), such that the $\tau_\mathrm{m.e.}$ value obtained from the fit was lying between them and 
could be associated with the ``mean'' maximal emission time with respect to the two maxima on the plot. 
The second maximum of 
the kaon emission function, causing the extracted maximal emission times for kaons be about $2-3$~fm/$c$ larger than 
those for pions, is likely connected with the decays of $K^*(892)$ resonance (its lifetime is about $4$~fm/$c$)
producing additional noticeable portion of kaons after particlization of the created fireball.

\begin{figure}
\centering
\includegraphics[width=0.8\textwidth]{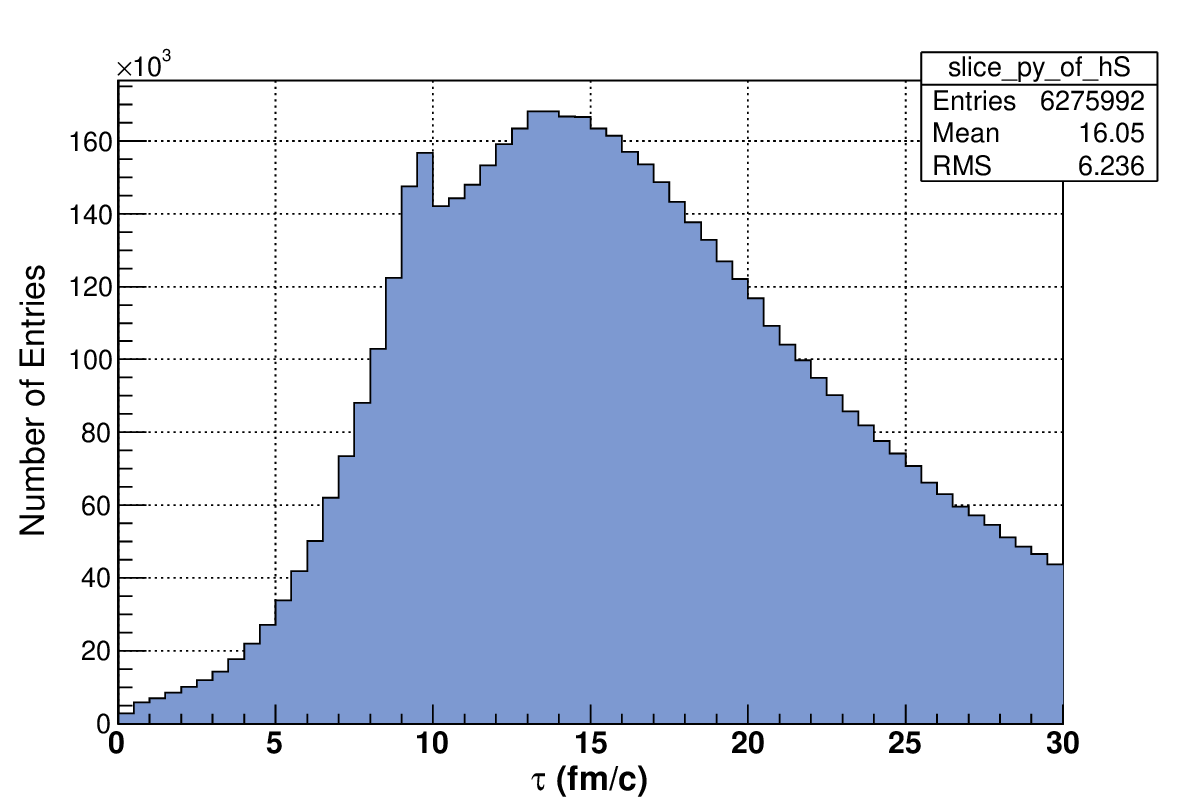}
\caption{The time of emission distribution for pions in central Pb+Pb collisions ($c=0-5\%$) at the LHC 
energy $\sqrt{s_{NN}}=2.76$~TeV simulated within iHKM, $0.2<p_T<0.3$~GeV/$c$, $|y|<0.5$.
\label{pismall}}
\end{figure}

\begin{figure}
\centering
\includegraphics[width=0.8\textwidth]{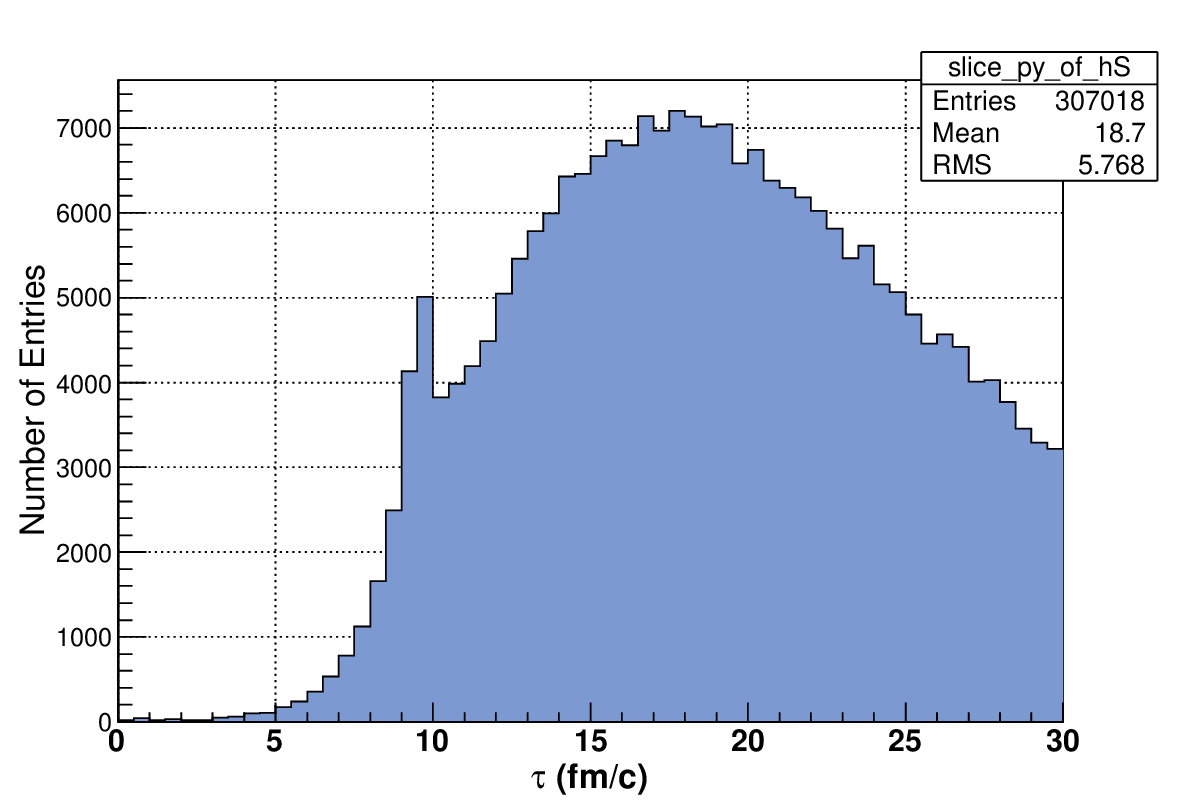}
\caption{The same as in Fig.~\ref{pismall} for kaons.
\label{ksmall}}
\end{figure}

To investigate the emission process in more detail we constructed the emission time distributions for both considered
particle sorts (see Figs.~\ref{pismall}, \ref{ksmall}), which can be obtained by integrating the previously analyzed 
emission function histograms over transverse radius $r_T$. An interesting thing one can notice about these
distributions is that the most particles of
both species seem to be produced at the late, hadronic stage of the collision, so that the total number of pions
and kaons emitted close to the hadronization hypersurface is noticeably smaller than that of mesons radiated
later. Near the time of hadronization in the center of the system ($9-10$~fm/$c$) one can see only local
$\tau$ distribution maxima on the presented plots. 

The found peculiarity of $\tau$ distributions is in agreement with our previous conclusions about
the continuous character of particle spectra freeze-out, however it is not clear why the extracted values 
of $\tau_\mathrm{m.e.}$, obtained from the combined fit to single-particle $p_T$ spectra and femtoscopy scales, 
do not reflect the apparent prevalence of the late post-hydrodynamic stage contribution in the total amount 
of emitted particles, following from the $\tau$ distribution figures.
 
Another interesting fact about the extracted maximal emission times is that their values are very close for 
the two considered LHC collision energies, $2.76$~TeV and $5.02$~TeV (the same applies to the times
of particlization of the corresponding systems). At first glance, it seems that
at higher collision energy the created system should live longer and produce more particles, so that the 
corresponding times of maximal emission should be larger. Instead at the two energies in our analysis 
for the most central collisions ($c=0-5\%$) we have comparable $\tau_\mathrm{m.e.}$ for both pions and kaons, 
$\tau_{\pi}=8.97 \pm 0.04$~fm/$c$ and $\tau_{K}=12.73 \pm 0.12$~fm/$c$ at $\sqrt{s_{NN}}=5.02$~TeV, 
and $\tau_{\pi}=10.34 \pm 0.06$~fm/$c$ (or even $\tau_{\pi}=9.44 \pm 0.02$~fm/$c$, 
see \cite{mtscale276} for details) and $\tau_{K}=12.65 \pm 1.58$~fm/$c$ at $\sqrt{s_{NN}}=2.76$~TeV.
The corresponding temperatures, entering the fitting formulas, are $T=138$~MeV and $T=144$~MeV respectively.

In the Section 4 we try to investigate in detail the stated above issues. 

\section{Materials and Methods}

As it was already mentioned, the analysis of relativistic heavy-ion collisions for this work is carried out based
on the results of computer simulations within the integrated hydrokinetic model 
(iHKM)~\cite{hkm,ihkm2016,ihkm2015,ihkm-review}. The model successfully reproduces
the experimental data on bulk hadronic observables (including femtoscopy radii) for the high-energy
collisions at the LHC and RHIC~\cite{ihkm2016,lhc502-ihkm,xenon,rhic-ihkm}. It also allows to achieve a good 
agreement with the data on direct photon production for the two mentioned collider 
experiments~\cite{ihkm1,ihkm2,ihkm-j}. 

One of the most important distinct features of iHKM,
as compared to other hybrid models used for the simulation of A+A collisions,
is the presence of the pre-thermal dynamics description for the strongly interacting matter
created as a result of collision (it is implemented using the energy-momentum
transport approach based on the Boltzmann equation in the relaxation time approximation ---
see the papers~\cite{ihkm2015,ihkm2016,ihkm-review} for details). 
During the corresponding evolution stage, the system is gradually 
transformed from an essentially non-equilibrium state it has at the very initial times right after 
the overlapping of the two colliding nuclei, to a nearly equilibrated state, which can be further described 
in the approximation of relativistic viscous hydrodynamics~\cite{IS3,hlle}. The initial
transverse energy-density distribution, 
serving a starting point for the pre-thermal stage, 
is generated with the GLISSANDO code~\cite{gliss}: 
\begin{equation}
\label{eps0}
\epsilon(b,\tau_0,\textbf{r}_{T})=\epsilon_{0}(\tau_0)
\frac{(1-\alpha)N_w(b,\textbf{r}_{T})/2+\alpha N_{bin}(b,\textbf{r}_{T})}{(1-\alpha)N_w(b=0,\textbf{r}_{T}=0)/2+\alpha N_{bin}(b=0,\textbf{r}_{T}=0)}.
\end{equation}
Here the parameter $\epsilon_0$ defines the initial energy density in the center of the system in very central collisions (when the parameter $b$ is small),
and the parameter $\alpha$ regulates the proportion between the contributions from the binary
collision and wounded nucleon models to the $\epsilon(b,\tau_0,\textbf{r}_{T})$ in (\ref{eps0}).
These two parameters are used for adjusting the model to the simulation of the concrete collision experiment.

For the initial momentum distribution (transverse and longitudinal) we take the anisotro\-pic momentum 
distribution, inspired by the Color Glass Condensate effective gluon field theory:
\begin{equation}
f_0(p)=g \exp\left(-\sqrt{\frac{(p\cdot U)^2-(p\cdot
V)^2}{\lambda_{\perp}^2}+\frac{(p\cdot
V)^2}{\lambda_{\parallel}^2}}\right),
\label{anis1}
\end{equation}
where $U^{\mu}=( \cosh\eta, 0, 0, \sinh\eta)$, $V^{\mu}=(\sinh\eta, 0, 0,\cosh\eta)$, $\eta$ is 
space-time longitudinal rapidity, and $\Lambda=\lambda_{\perp}/\lambda_{\parallel}=100$ 
is the model parameter, describing the initial momentum anisotropy 
(see~\cite{ihkm2015,ihkm2016,ihkm-review} for details).

The hydrodynamic stage of the matter expansion lasts until
the temperature in the system drops to the particlization value $T_p$ (depending on the 
equation of state (EoS) utilized for the continuous quark-gluon matter; in iHKM we compare 
the results obtained using the Laine-Schroeder~\cite{laine} and the HotQCD Collaboration~\cite{hotqcd} EoSs). 
Below the particlization temperature,
the continuous medium transforms into the system of hadrons, which keeps on expanding, while the hadrons
interact intensively with each other, so that numerous particle creation and annihilation processes take place. 
This post-hydrodynamic stage of the matter evolution is described with the help of the UrQMD hadron cascade 
model~\cite{urqmd1,urqmd2}.

The analytical formulas for spectra and femtoscopy radii fitting in~\cite{mtscale276, rhic-ihkm, mtscale502}
are obtained from the following considerations. 

One assumes that the particlization occurs at the hypersurface $\tau=const=\tau_\mathrm{m.e.}$,
limited in the direction transverse to the beam axis ($r_T \le r^\mathrm{max}_T$),
which should be reasonable at least for the particles with not very high momenta ($p_T \lesssim 0.5$~GeV/$c$). 
Accordingly, the Wigner function for soft bosons can be written as follows
\begin{equation}
f_{l.eq.}(x,p)=\frac{1}{(2\pi)^3}\left[\exp(\beta p\cdot u(\tau_{m.e.},{\bf r}_T) -\beta\mu)-1\right]^{-1}\rho({\bf r}_T),
\label{wigner}
\end{equation} 
where $\beta$ is the inverse temperature, 
$u^{\mu}(x)=(\cosh\eta_L\cosh\eta_T,\frac{{\bf r}_T}{r_T}\sinh\eta_T,\sinh\eta_L\cosh\eta_T)$ is 
the collective 4-velocity (here in $u^{\mu}$ definition $\eta_L= \text{arctanh}\,v_L$ 
and $\eta_T = \text{arctanh}\,v_T(r_T)$ are longitudinal and transverse rapidities,
which correspond to respective velocities $v_L$ and $v_T$), 
and $\rho({\bf r}_T)$ is the cutoff factor, narrowing the particlization
hypersurface in the transverse direction, 
\begin{equation}
\rho({\bf r}_T)=\exp[-\alpha (\cosh\eta_T(r_T)-1)].
\label{rho}
\end{equation}
The parameter $\alpha$ here describes the transverse collective flow intensity, in such a way
that the lower is the value of $\alpha$, the stronger is the flow 
(in case of absent flow $\alpha\rightarrow \infty$; for more details see~\cite{mtscale276, mtscale502}).   

Given the Eq.~(\ref{wigner}) for the Wigner function, one can obtain
the approximate formulas for the momentum spectra and correlation functions.
Applying the modified Cooper-Frye prescription (it suggests that overall particlization
hypersurface is built as a set of points $(t(\textbf{r}, p); \textbf{r})$, 
corresponding to the maximal emission of particles with momentum $p$, see~\cite{hkm} for details) 
at the previously defined particlization hypersurface
and using the saddle-point method, one arrives at the following expressions
for the $p_T$ spectra and the longitudinal femtoscopy radii~\cite{tolstykh}:
\begin{equation}
p_0 \frac{d^3N}{d^3p} \propto \exp{[-(m_T/T + \alpha)(1-\bar{v}^2_T)^{1/2}]},
\label{specfit}
\end{equation}
\begin{equation}
R^2_{\mathrm{long}}(m_T)=\tau^2\lambda^2\left(1+\frac{3}{2}\lambda^2\right).
\label{radfit}
\end{equation}
Here $m_T=\sqrt{m^2+k_T^2}$, $k_T$ is the pair mean transverse momentum,
$T=T_\mathrm{m. e.}$ is the temperature at the maximal emission hypersurface $\tau=\tau_\mathrm{m.e.}$,
$\bar{v}_T=k_T/(m_T+\alpha T)$ is the transverse collective velocity at the saddle point, and
\begin{equation}
\lambda^2 = \frac{T}{m_T}(1-\bar{v}^2_T)^{1/2}
\label{lambda}
\end{equation}
is the squared ratio of the longitudinal homogeneity length $\lambda_\mathrm{long}$ to $\tau_\mathrm{m.e.}$
(as usual in femtoscopy studies, $long$ direction coincides with that of the beam axis,
$out$ corresponds to the pair transverse momentum vector, and $side$ is orthogonal to the two others).
The Eq.~(\ref{radfit}) is valid for the transverse flow of arbitrary profile and strength,
however it was derived under assumption of small $q_\mathrm{long}$, corresponding to the top part of the 
correlation function peak.

According to the method, originally proposed in~\cite{mtscale276}, in order to estimate the $\tau_\mathrm{m.e.}$
values for pions and kaons, one needs to find the values of $T$ and $\alpha$ first from the combined fit 
to the pion and kaon $p_T$ spectra based on Eq.~(\ref{specfit}), and after that determine the maximal emission times 
$\tau_{\pi}$ and $\tau_{K}$ from the femtoscopy radii $R_{\mathrm{long}}(m_T)$ fitting 
(at already fixed $T$ and $\alpha$) using Eq.~(\ref{radfit}).

\section{Results and Discussion}

\subsection{Emission time distributions}

As it was already mentioned, the emission time distributions for both pions and kaons,
shown in Figs.~\ref{pismall}, \ref{ksmall} for the case of $2.76$~TeV Pb+Pb central collisions
at the LHC, suggest that the large number of particles
leave the system noticeably later, than the time of full particlization, which is about 9-10~fm/$c$
for the considered LHC energies (rather similar plots are obtained for $5.02$~TeV collisions as well).  
The mean $\tau$ values that can be calculated based on
the presented model histograms are 16.05 fm/$c$ and 18.70 fm/$c$ for pions and kaons respectively,
and an interesting question is why these time scales do not match with $\tau_\mathrm{m.e.}$ estimated
based on the femtoscopy radii $m_T$ dependencies (i.e., 10.3 fm/$c$ and 12.7 fm/$c$).

Here one should consider several different factors, that could possibly lead to such a result, 
and try to find out which of them played the major role in this situation.

\begin{figure}
\centering
\includegraphics[width=0.8\textwidth]{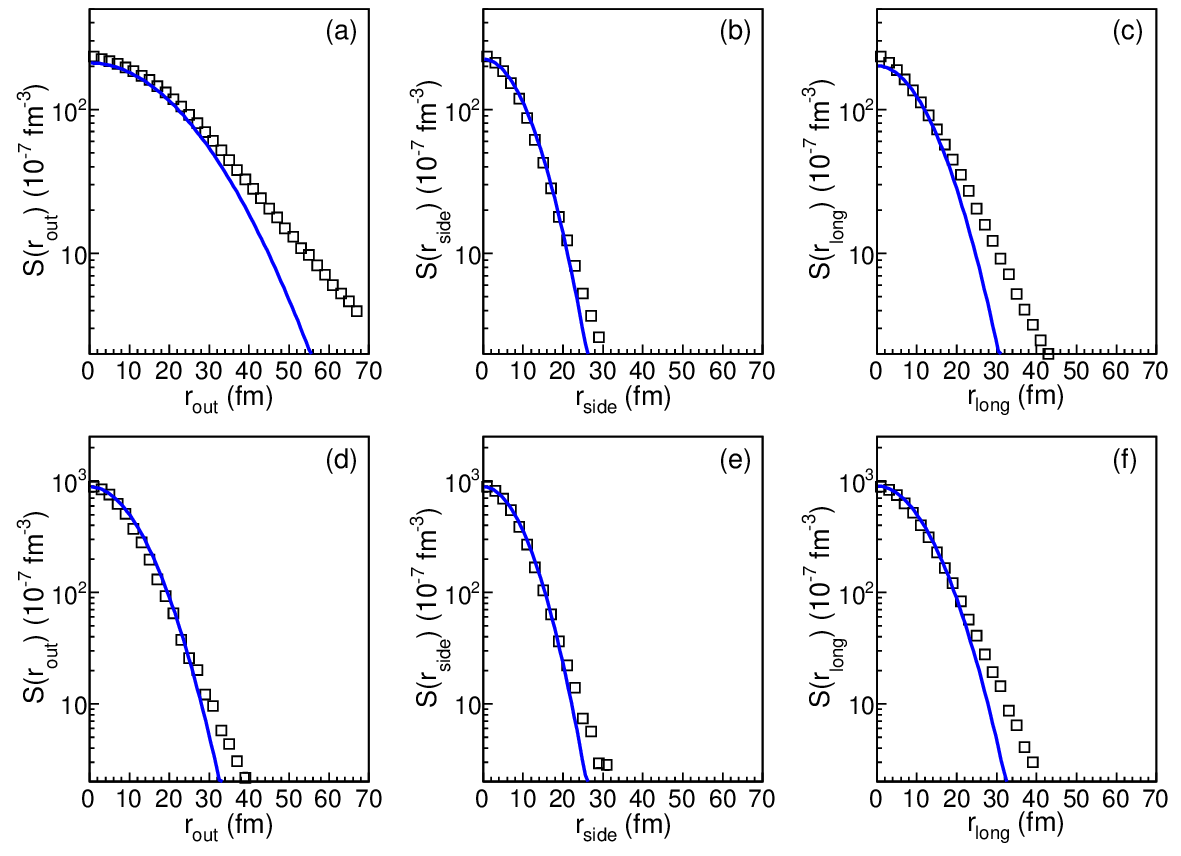}
\caption{The source function projections 
calculated in iHKM model for pions (panels (a), (b), (c)) and kaons (panels (d), (e), (f))
produced in central Pb+Pb collisions at the LHC energy $\sqrt{s_{NN}}=2.76$~TeV, $0.2 < p_T < 0.36$~GeV/$c$, $|y|<0.5$. 
Squares represent the model output and the lines show the corresponding Gaussian fits.
\label{sflhc}}
\end{figure}

One of the relevant factors could be the non-Gaussian shape of the corresponding particle emission
functions and the correlation functions, which are used to determine the femtoscopy radii.
In \cite{sf} the non-Gaussian character of pion and kaon emission was observed in our studies of 
the corresponding source functions (which can be defined as time-integrated distributions 
of distances between the particles forming pairs in the pair rest frame), calculated in the HKM model
--- a previous version of iHKM --- for the case of Pb+Pb collisions at the LHC 
energy $\sqrt{s_{NN}}=2.76$~TeV.  
In Fig.~\ref{sflhc} we demonstrate the projections of the mentioned meson source functions together
with the Gaussian fits to them. As one can see, the actual shape of the source functions,
especially in \textit{out} and \textit{long} directions, includes noticeable non-Gaussian tails.
The effect looks even more pronounced for pions.  
The femtoscopy radii are extracted from the Gaussian fits to the correlation 
functions $C(\textbf{q})$, which are connected with the source functions $S(\textbf{r})$ through 
the so-called Koonin integral equation~\cite{koonin}
\begin{equation}
C(\textbf{q}^{*})=1+ \int d^3r^{*} S(\textbf{r}^{*}) K(\textbf{r}^{*},\textbf{q}^{*}),
\label{koon}
\end{equation}
where $K(\textbf{r}^{*},\textbf{q}^{*})$ is the integral transform kernel, reflecting the 
correlation mechanism(s) supposed to exist between the emitted particles, and asterisk denotes
the pair rest frame. 

Accordingly, the pairs from the non-Gaussian tail of the source function,
characterized by the \textit{large distances} between emitted particles will not contribute directly
to the value of the femtoscopy radius, and, eventually, to the $\tau_\mathrm{m.e.}$ 
value~\footnote{As it follows from Eq.~(\ref{radfit}), the larger is the $R_{long}$ value, the larger 
can be the estimated $\tau_\mathrm{m.e.}$.}.
In such a way some amount of particles which escape from the system at the late stage
of the collision can be missed by the femtoscopy analysis, but still be visible in
overall $\tau$ distribution. 

In correlation function plots, such pairs, containing,
e.g. particles coming from the long-lived resonance decays or liberated from the system
after a series of elastic/inelastic reactions with other hadrons constituting the expanding
hadronic medium (so that the distance between the two particles in pair can be fairly large),
typically form a practically invisible sharp and narrow peak~\cite{schlei2} 
which can reduce the intercept $\lambda$ value,
determined from the fit to the correlation function, but does not really affect the radii.
  
In \cite{mtscale276} we tried to reduce the effects stemming from non-gaussianity of the correlation
functions by fitting them in a narrow $q$ range ($0<q_\mathrm{long}<0.04$~GeV/$c$), which also better
corresponds to the assumptions made when deriving formula (\ref{radfit}). As a result, we observed
more uniform behavior of pion and kaon radii, closer to $m_T$ scaling (still broken because of
collective flow and afterburner rescattering effects), with common $T$ and $\alpha$ parameters
corresponding to the combined pion and kaon $p_T$ spectra fitting, however
only a hardly noticeable increase of the femtoscopy radii (about 4\%) could be reached. 

The presence of a post-hydrodynamic hadronic cascade stage in the model certainly increases
the resulting femtoscopy radii --- in our analysis carried out during the work on the papers~\cite{klhc,mtscale276}
we found for the case of the LHC $2.76$~TeV Pb+Pb collisions, that the switching off the cascade
stage reduces the radii values by 12\% for pions and by 25-30\% for kaons.
However the maximal emission time extraction we are currently interested in is performed based on the full
calculation mode results, i.e. with the hadron cascade turned on, so the above observation can hardly help us to solve the stated problem.  

\begin{figure}
\centering
\includegraphics[width=0.8\textwidth]{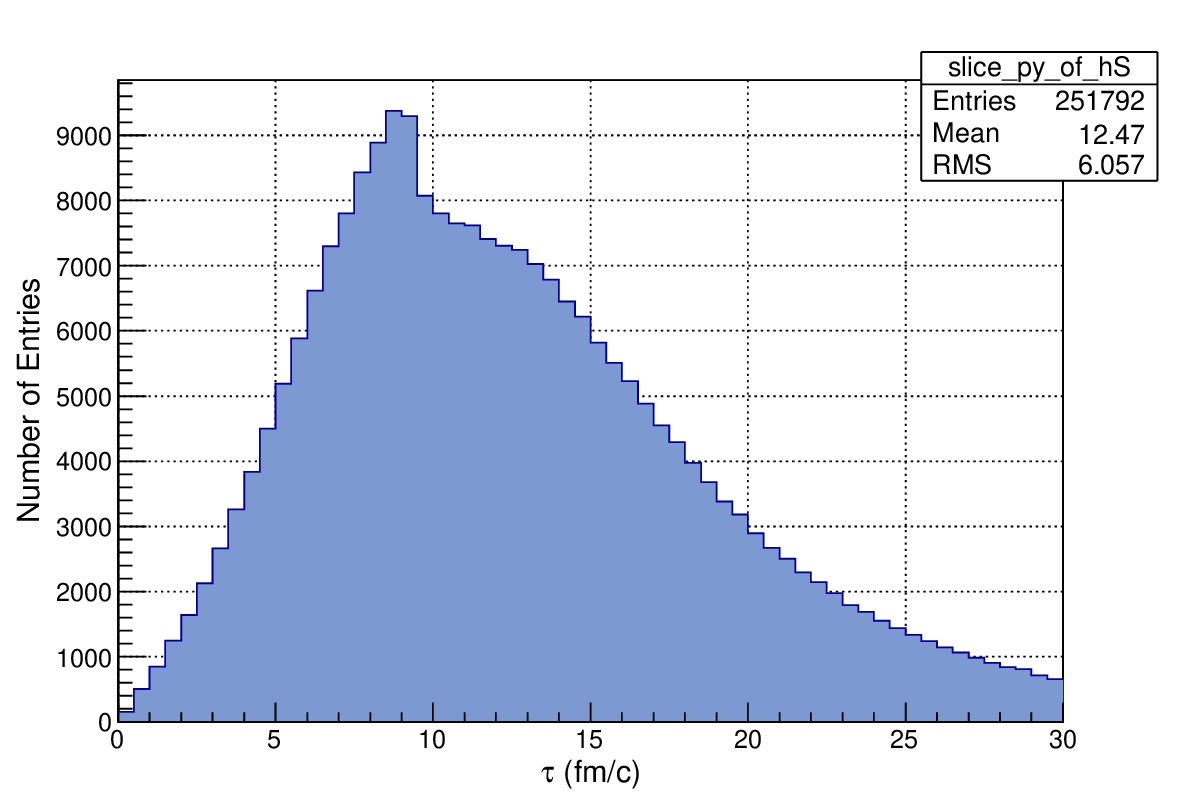}
\caption{The same as in Fig.~\ref{pismall}, but for wide $p_T$ 
region, $0.5<p_T<2.0$~GeV/$c$.
\label{piwide}}
\end{figure}

\begin{figure}
\centering
\includegraphics[width=0.8\textwidth]{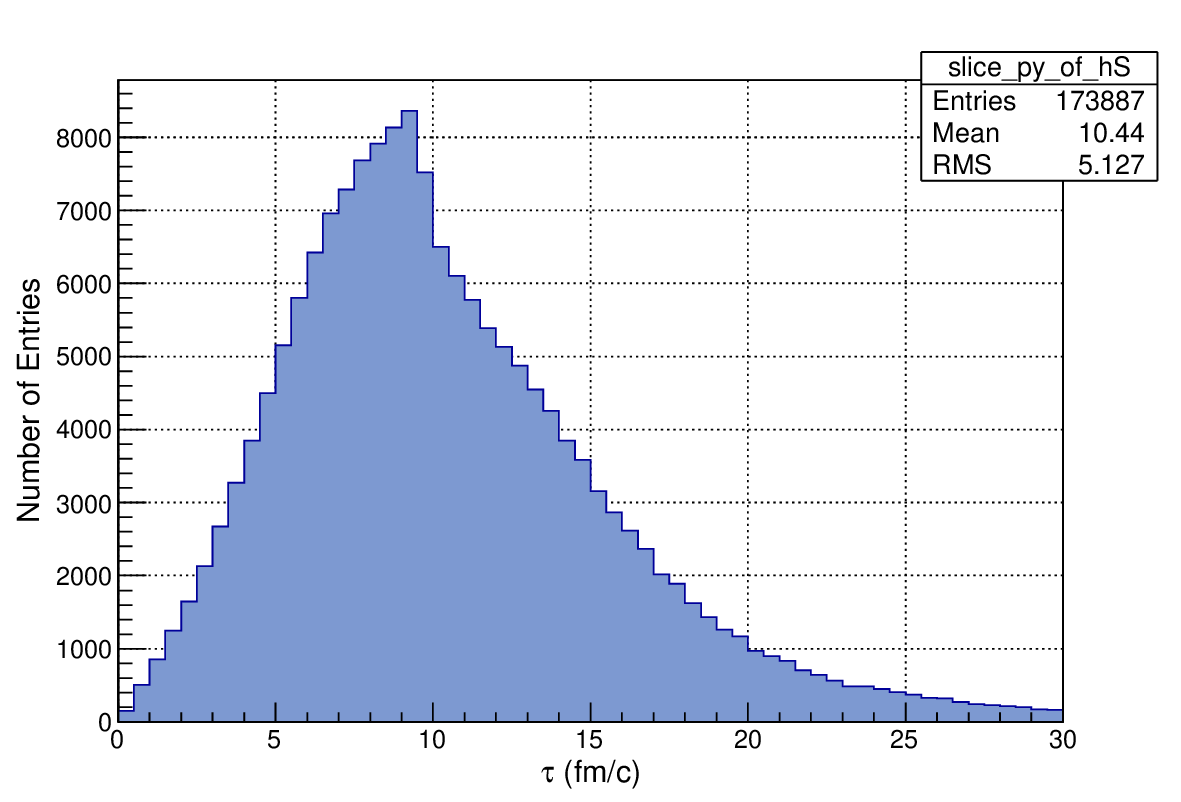}
\caption{The same as in Fig.~\ref{piwide}, but with the constraint $r_T<10$~fm.
\label{piwidert}}
\end{figure}

A more thorough analysis of the procedure we apply for the $\tau_\mathrm{m.e.}$ extraction, however,
suggests the following possible solution. Although the analytical formulas, (\ref{specfit}) and (\ref{radfit}), 
used for the spectra and the femtoscopy radii fitting were derived under assumption of isochronous
particle emission from the hypersurface $\tau=const=\tau_\mathrm{m.e.}$ fragment with $r_T<r_T^\mathrm{max}$, 
typical for particles with not very high transverse momenta, the actual data fitting included also the region of
comparably high $p_T$ ($0.5<p_T<1.0$~GeV/$c$) and $k_T$ ($0.3<k_T<1.1$~GeV/$c$). 
This apparently implies that the emission time
distributions, actually corresponding to the applied fitting scheme, should as well include particles
from a wide momentum region (still not exceeding, however, the region of hydrodynamics approximation applicability).
Therefore, in Figs.~\ref{piwide}, \ref{piwidert} we show the pion $\tau$ distributions constructed 
with a momentum cut $0.5<p_T<2.0$~GeV/$c$ for central Pb+Pb collisions at the LHC energy $\sqrt{s_{NN}}=2.76$~TeV (analogous distributions for kaons demonstrate similar tendency as compared to the previous Fig.~\ref{ksmall}). 
In Fig.~\ref{piwidert} additional $r_T$ constraint, $r_T<10$~fm, was applied.

As one can see, the presented histograms correspond much better to the $\tau_\mathrm{m.e.}$ estimates
we previously obtained from the spectra and radii fitting, so that the distribution maxima are close to the full particlization
time ($\approx 10$~fm/$c$), the tails corresponding to the hadronic phase look much smaller and 
do not dominate in the overall distributions, and the mean emission time values are 12.5~fm/$c$ and 
10.4~fm/$c$ for $r_T$-unconstrained and $r_T$-constrained cases respectively.

The obtained wide-momentum-range $\tau$ distributions seem to be more relevant for characterizing
the emission process from the system as a whole. But what about the low-momentum distributions shown before 
in Figs.~\ref{pismall}, \ref{ksmall} and depicting a quite rich and prolonged hadronic stage of the matter
evolution? Could we use these distributions for getting estimates of the afterburner phase duration, 
corresponding to the ``maximal emission times'', but now determined based on low-momentum spectra and radii?
In principle, such an estimate would be in agreement with the modified Cooper-Frye prescription~\cite{hkm},
used in~\cite{mtscale276,mtscale502} for the derivation of the formulas~(\ref{specfit}), (\ref{radfit})
and suggesting that the particles with different momenta $p$ should be emitted from separate pieces $\sigma_p$ 
of the complex hadronization hypersurface, composed as a united set of all such fragments.

So, to test such an approach we used the described method for the $\tau_\mathrm{m.e.}$ estimation, 
but applying the low-momentum cuts on $p_T$ and $k_T$. 
The combined $\pi$ and $K$ spectra fitting was performed for the $p_T$ range $0.25<p_T<0.55$~GeV/$c$
and gave the values $T=106\pm16$~MeV, $\alpha_{\pi}=2.8\pm0.5$, $\alpha_{K}=1.2\pm0.7$, which were then fixed at 
the $R_\mathrm{long}(m_T)$ fitting.
The latter was done for $k_T<0.6$~GeV/$c$ and resulted in the maximal emission time values 
$\tau_{\pi}=13.65\pm0.44$~fm/$c$ and $\tau_{K}=16.73\pm0.49$~fm/$c$. 

\begin{figure}
\centering
\includegraphics[width=0.98\textwidth]{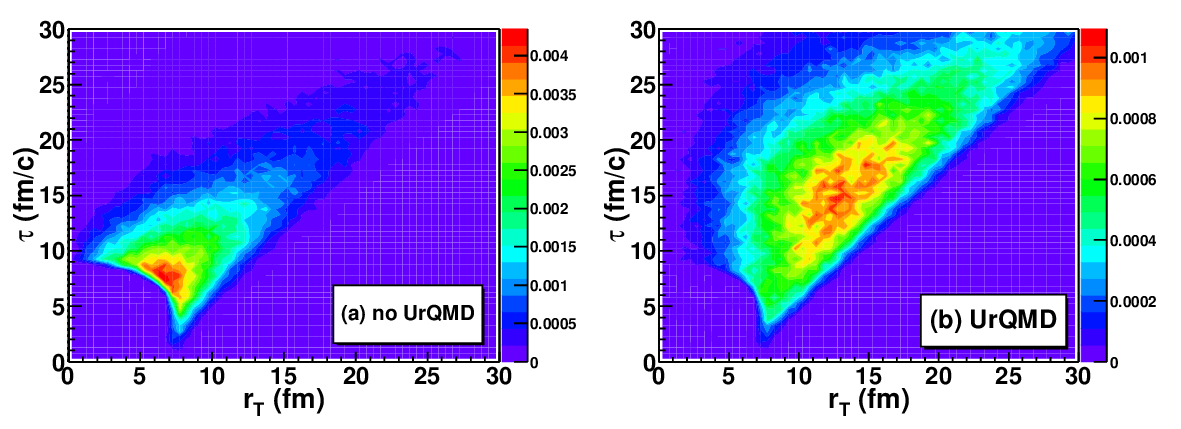}
\caption{Emission functions $g(\tau, r_T)$ constructed in iHKM for $K^{+}\pi^{-}$ pairs,
coming from $K^*(892)$ decays~\cite{kstar1}, in two regimes: 
(a) free streaming of hadrons created at the particlization stage (plus those coming from resonance decays),
and (b) full iHKM calculation including UrQMD cascade as the final afterburner stage.
The graphs correspond to the LHC Pb+Pb collisions at $\sqrt{s_{NN}}=2.76$~TeV, $c=5-10\%$,
the transverse momentum and rapidity ranges are $0.3<k_T<5$~GeV/$c$ and $|y|<0.5$ respectively. 
\label{kstaremiss}}
\end{figure}

Comparing these times with
the Figs.~\ref{pismall}, \ref{ksmall} one can see that the new $\tau_\mathrm{m.e.}$ values correspond
to the maxima of the shown low-momentum emission time distributions and
can be possibly interpreted as an evidence of prolonged afterburner stage and used as ``upper-limit'' estimates
for the overall system's lifetime (in the sense that they are defined based on low-momentum data,
and particles with low $p_T$ are expected to leave the system later than those with high $p_T$).
The new $\tau_\mathrm{m.e.}$ value for kaons is also in agreement
with the emission picture presented in Fig.~2 of our paper~\cite{kstar1}, 
which implies that active $K^*$ production at the hadronic
stage of collision lasts until $\approx 15-20$~fm/$c$, while $K^*$ resonance decays serve as an extra 
source of kaons (see also Fig.~\ref{kstaremiss} of this article).

\subsection{Particlization times at different collision energies}

In previous subsection we focused on the particle emission picture including long
and intensive emission from the hadronic stage of collision and on estimation of the effective time
scales characterizing the emission duration. However, a detailed analysis of this problem
cannot bypass the question of the hydrodynamic stage duration, defining the moment, when one should switch 
from the description of the system's evolution in terms of continuous medium to the description
in terms of particles. 

Conventionally, one assumes that the system's hadronization takes place at some critical temperature
or energy density (whose concrete values depend on the applied equation of state for quark-gluon matter).
As it was mentioned, in iHKM we use two such equations of state, namely the Laine-Schroeder~\cite{laine} EoS 
and the HotQCD Collaboration~\cite{hotqcd} one. The former is associated with the particlization temperature
$T_p=165$~MeV and the corresponding energy density $\epsilon_p=0.5$~GeV/fm$^3$, while for the latter
the values $T_p=156$~MeV and $\epsilon_p=0.27$~GeV/fm$^3$ are taken. The stated switching criterion
implies that the particlization process is not isochronous for different parts of the system, but starts 
quite early at the periphery and gradually reaches the center. Thus, the system can be considered
fully hadronized when the hadronization has completed in its center.

As we already remarked in the Introduction, an interesting fact about the particlization times in
ultrarelativistic heavy ion collisions at the top RHIC and the LHC energies is that in iHKM they are very close
to each other (about $8-10$~fm/$c$), despite great differences in the collision energies (and, apparently, the 
corresponding initial energy densities). This feature is also reflected in the close maximal emission times
obtained for pions in our studies related to the corresponding 
collision experiments~\cite{mtscale276,rhic-ihkm,mtscale502}.

\begin{figure}
\centering
\includegraphics[width=0.8\textwidth]{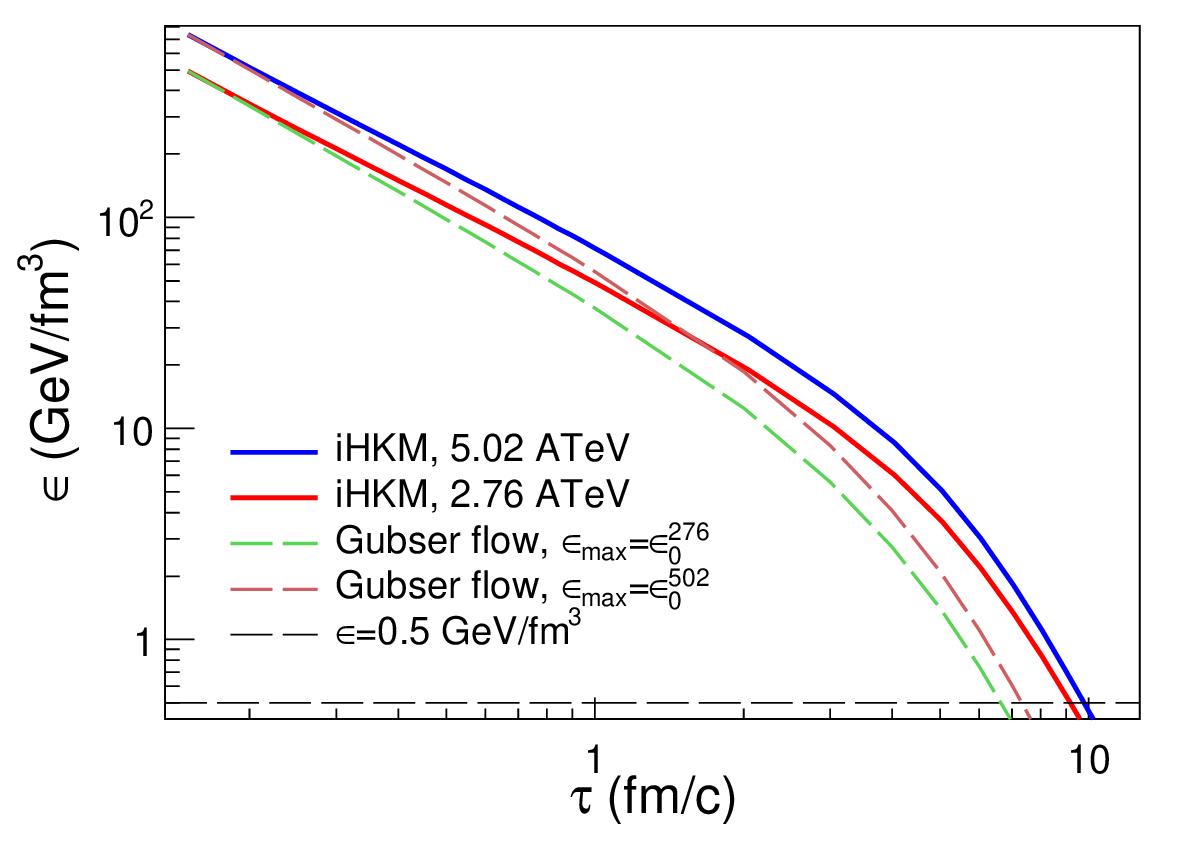}
\caption{The energy density $\epsilon$ at the center of the system depending on proper time $\tau$. 
The iHKM curves for the two LHC energies,
$\sqrt{s_{NN}}=2.76$~TeV and $\sqrt{s_{NN}}=5.02$~TeV, together with the 3D Gubser solutions (\ref{gubser}) of
relativistic hydrodynamics equations. The parameter $q$ value, $q=0.15$~fm$^{-1}$,
is defined from the Gubser fit to the initial iHKM $\epsilon(r_T)$ profiles, the initial $\epsilon$ values
correspond to the iHKM ones.
\label{epst1}}
\end{figure}

\begin{figure}
\centering
\includegraphics[width=0.8\textwidth]{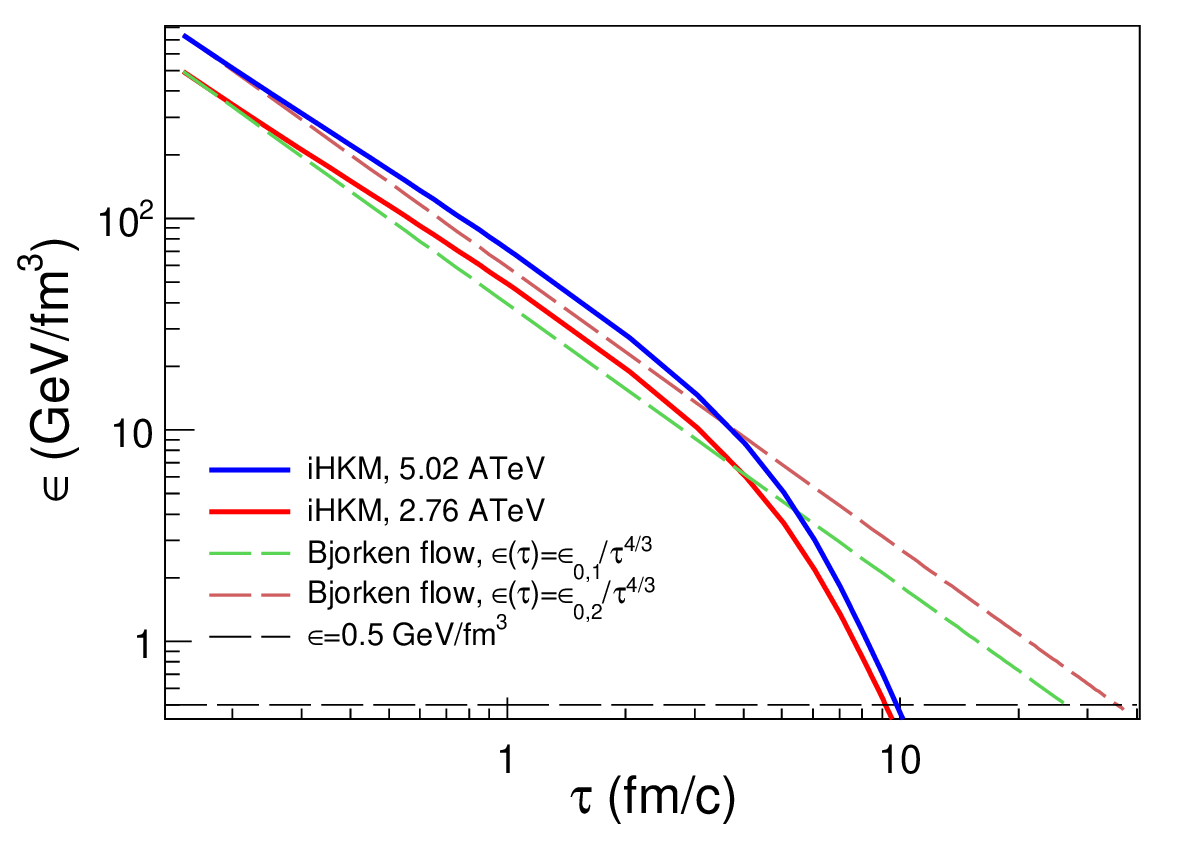}
\caption{The same as in Fig.~\ref{epst1}, but 1D Bjorken solutions, $\epsilon(\tau)=\epsilon_0 \tau^{-4/3}$,
are shown for comparison with the iHKM curves.
\label{epst2}}
\end{figure}

To examine this issue we built plots of the time evolution for the energy density in the 
center of hydrodynamic grid $\epsilon(\tau)$ within iHKM (see Figs.~\ref{epst1}, \ref{epst2})
in case of the LHC Pb+Pb collisions at the two energies, $\sqrt{s_{NN}}=2.76$~TeV and $\sqrt{s_{NN}}=5.02$~TeV.
The Laine-Schroeder EoS was used in simulations for both energies with the critical energy density
$\epsilon_p=0.5$~GeV/fm$^3$.
For comparison we also placed the curves corresponding to the ideal relativistic hydrodynamics 
three-dimensional Gubser solution~\cite{gubs},
\begin{equation}
\epsilon(r_T,\tau)= \frac{\epsilon_0}{\tau^{4/3}}\frac{(2 q)^{8/3}}{[1+2q^2(\tau^2+r_T^2)+q^4(\tau^2-r_T^2)^2]^{4/3}},
\label{gubser}
\end{equation}
and Bjorken solution $\epsilon(\tau)=\epsilon_0 \tau^{-4/3}$ for purely longitudinal
expansion, in the presented figures. The initial maximal $\epsilon$ values for the Gubser and the Bjorken flows
correspond to those in iHKM. Also, in the Gubser flow case, the parameter $q$ value, $q=0.15$~fm$^{-1}$,
defining the transverse width of the energy density profile was chosen based on the fit to
the initial iHKM $\epsilon(r_T)$ profile.

From the comparison with the 1D Bjorken flow case (see Fig.~\ref{epst2}) one can conclude that the presence of strong
transverse flow along with the longitudinal one is an important factor leading to the observed
small difference in particlization times for the two collision energies (about 9.2~fm/$c$ and 9.8~fm/$c$
for iHKM in the shown figure), defined as the moments when the respective energy densities 
drop to $\epsilon_p=0.5$~GeV/fm$^3$. Indeed, we see that at the times after 3~fm/$c$ Bjorken
expansion is much slower than in iHKM, the ratio between the two $\epsilon(\tau)$ (about 1.5)
does not change with time, so that the respective particlization times (about 26~fm/$c$ and 36~fm/$c$) 
are much higher than in our model, the ratio between these times, $(\epsilon_{0,2}/\epsilon_{0,1})^{3/4}$, 
is about 1.35, and therefore the difference between them is quite large (about 10~fm/$c$).  

The Gubser flow, accounting for the transverse matter expansion, results in an $\epsilon(\tau)$
behavior more similar to the iHKM curves at $\tau>3$~fm/$c$ and particlization times
much closer to those from iHKM (about 6.7~fm/$c$ and 7.3~fm/$c$), also with a small difference
between the two values. The analysis of the particlization time $\tau_p$ dependency on the initial energy
density $\epsilon_0(\tau_0)$ in iHKM shows that for the considered RHIC and LHC energy range in central collisions 
at fixed EoS and initial time $\tau_0$, $\tau_p$ in iHKM grows slowly with $\epsilon_0$,
approximately as $\epsilon_0^{1/s}$, where $s$ is close to 7. In our opinion, such a behavior
can be explained by the intensive 3D expansion of the system, which takes place at large $\tau$
in high-energy A+A collisions and leads to a fast subsequent decay of the created fireball.
The suggestion about the breakdown of hydrodynamic description as a result of intensive 
system's 3D expansion for the systems formed in high-energy hadron collisions was made already in 1953 
by Landau in his pioneer paper concerning hydrodynamic approach in high-energy physics~\cite{landau}.

Energy density behavior, somewhat similar to that shown in Figs.~\ref{epst1}, \ref{epst2} can be also observed 
in the following non-relativistic analytical
solution of both ideal hydrodynamics and Boltzmann equations (see, e.g.~\cite{hydrosol,freeze}). 
The corresponding distribution function can be written as
\begin{equation}
f(t,\textbf{x},\textbf{v})= \frac{N}{(2 \pi R_0)^{3}} \left(\frac{m}{T_0}\right)^{3/2}\exp{\left(-\frac{m\textbf{v}^2}{2T_0}-\frac{(\textbf{x}-\textbf{v}t)^2}{2R_0^2}\right)},
\label{distf}
\end{equation}
where $N$ is total particle number, $m$ is particle mass, $R_0$ is the initial Gaussian radius of the system, 
$T_0$ is the initial temperature.

Given the distribution function (\ref{distf}), one easily finds the corresponding
dependency of the energy density at $\textbf{x}=0$ on time:
\begin{equation}
\epsilon(\textbf{x}=0,t)= \frac{A T_0^4 (m R_0^2/T_0)^{5/2}}{(t^2+m R_0^2/T_0)^{5/2}},
\label{solut}
\end{equation}
where $A$ is a constant.

\begin{figure}
\centering
\includegraphics[width=0.8\textwidth]{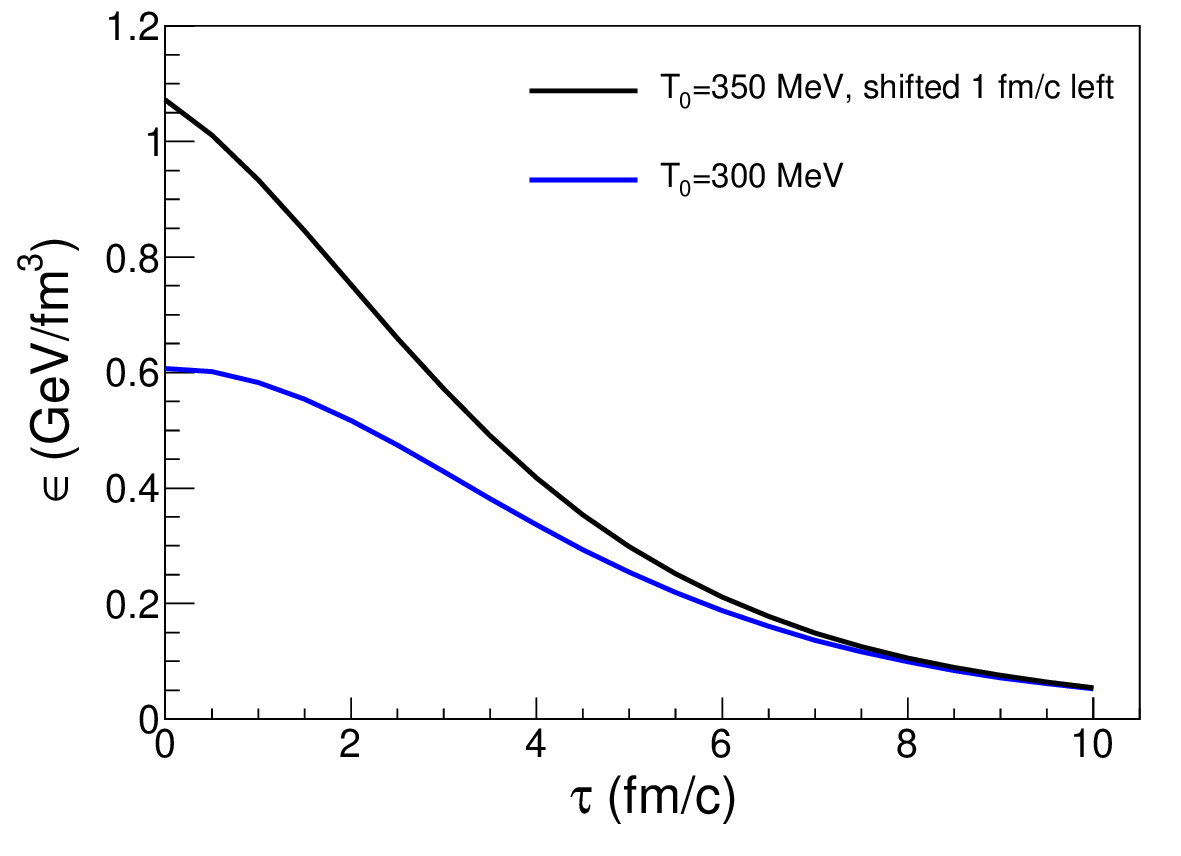}
\caption{The energy density $\epsilon$ at the center of the system dependencies on time $\tau$ 
in non-relativistic hydrodynamics and Boltzmann equations solution (\ref{solut}) with the two different
initial $\epsilon$ values reaching the value $\epsilon=0.5$~GeV/fm$^3$ almost simultaneously (see text for details).
\label{epstnr}}
\end{figure}

In Fig.~\ref{epstnr} we present the illustration corresponding to the solution (\ref{solut}).
The black line corresponds to a higher initial energy density, regulated by the $T_0$ parameter,
which in this case has the value $T_0=350$~MeV. For the blue line $T_0=300$~MeV, and the initial
energy density is about 2 times lower than for the black line. In both cases, the Gaussian radius of the
initial fireball is $R_0=6$~fm, and mass of the particles in the gas is $m=0.5$~GeV/$c^2$.
The black curve in the figure is shifted 1 fm/$c$ left, so that after $t=8$~fm/$c$ we have the same energy
density for both evolving systems. So, similarly to the Figs.~\ref{epst1}, \ref{epst2}, the same ``particlization''
conditions are reached at nearly the same expansion times for the systems with the two different initial 
energy densities/temperatures, when the system's expansion becomes essentially 3-dimensional.

\section{Conclusions}

We considered the two paradoxical femtoscopic observations in ultrarelativistic heavy ion collisions. 
The first one is closeness of the observed maximal emission times at the quite different collision energies. 
It can be explained by the \textit{intensive} 3D (!) expansion of the system, which leads, starting from some time
to a fast decay of the formed continuous medium for a wide range of high collision energies.
Therefore, all the corresponding systems reach the decay energy almost simultaneously. 

Another paradoxical effect is that despite the long enough duration of the post-hydro\-dynamic/afterburner cascade 
stage, the observed times of the maximal emission are close to the particlization times.
The key to the answer lies in different pictures of particle radiation in the narrow soft $p_T$ range and in 
a more wide one, typical for the femtoscopy analysis. For the latter the times of maximal emission extracted 
from the corresponding fits nearly coincide with the particlization times. For the former, times are significantly 
larger and reflect the duration of the afterburner stage at this momentum interval. 

\acknowledgments{The research was carried out within the NAS of Ukraine Targeted Research
Program ``Collaboration in advanced international projects on high-energy physics and 
nuclear physics'', Agreement No.~7/2023 between the NAS of Ukraine and BITP of the NAS
of Ukraine.
The work was also supported by a grant from the Simons Foundation 
(Grant Number 1039151, Yu.S., V.S., M.A.).
The author Yu.S. is supported by the Excellence Initiative Research University grant 
of the Warsaw University of Technology.

Yu. S. is grateful to J. Schukraft, 
J. Stachel, P. Braun-Munzinger, and A. Dainese for stimulating and fruitful discussions.
Yu. S. is also grateful to ExtreMe Matter Institute (EMMI) at GSI in Darmstadt, 
where he was affiliated at the beginning of the work on the current paper, and to Hanna Zbroszczyk 
for her invitation to continue these studies in Warsaw University of Technology (WUT) as the visiting professor.}


\end{document}